\documentclass[12pt]{article}

\usepackage{url}
\usepackage{multirow}
\usepackage{lmodern} % readable font on Windows
\usepackage{color,soul}
\usepackage{graphicx}
\usepackage{subfig} % For subfigures
\usepackage{lineno} % For numbered lines
\usepackage{authblk} % Manages authors' affiliation
\usepackage{amsmath} % Math symbols
\usepackage{bm} % Bold math symbol
\usepackage{amssymb} % For real number like symbols
\usepackage[superscript]{cite} % citations with superscript.
\usepackage[inline]{enumitem} % For lists in the text
\usepackage[textsize=scriptsize]{todonotes} % To add comments
\usepackage[labelsep=period]{caption} % "Figure 1." instead of "Figure 1:"
\usepackage{arydshln} %dashed line in table \hdashline
\usepackage[utf8]{inputenc} % For accented characters
\usepackage[T1]{fontenc} % For accented characters
\usepackage[font=scriptsize]{caption} % To reduce caption font size
\reversemarginpar % Moves the todo comments on the left side

\DeclareMathOperator*{\argmin}{argmin}  
\DeclareMathOperator*{\argmax}{argmax}  
\DeclareMathOperator{\diag}{diag}  
\usepackage{comment}

\def\*#1{\mathbf{#1}} % instead of using \mathbf use \*

\begin{document}

\title{Functional clustering methods for resistance spot welding process data in the automotive industry}

\author{Christian Capezza}
\author{Fabio Centofanti}
\author{Antonio Lepore}
\author{Biagio Palumbo\thanks{Corresponding author e-mail: \texttt{biagio.palumbo@unina.it}}}
\affil{Department of Industrial Engineering, University of Naples Federico II, \\ Piazzale Vincenzo Tecchio 80, 80125, Naples, Italy}
\renewcommand\Affilfont{\itshape\small}
\date{}
\maketitle

%\pagenumbering{gobble}
%\clearpage
%\pagenumbering{arabic}

%\begin{center}
%\LARGE
%Control charts for monitoring ship operating conditions and CO\textsubscript{2} emissions based on scalar-on-function regression
%\end{center}

\normalsize

\begin{abstract}
%MAX 246/250 words
%1. An introduction to the work. This should be accessible by scientists in any field and express the necessity of the experiments executed
%2. Some scientific detail regarding the background to the problem
%3. A summary of the main result
%4. The implications of the result
%5. A broader perspective of the results, once again understandable across scientific disciplines
%It is crucial that the abstract conveys the importance and novelty of the work and be understandable without reference to the rest of the manuscript to a multidisciplinary audience. Abstracts should not contain any citation to other published works.

Quality assessment of resistance spot welding (RSW) joints of metal sheets in the automotive industry is typically based on costly and lengthy off-line tests that are unfeasible  on the full production, especially on large scale. However, the massive industrial digitalization triggered by the  industry 4.0 framework makes available, for every produced joint, on-line RSW process parameters, such as, in particular, the so-called dynamic resistance curve (DRC), which is recognized as the full technological signature of the spot welds.
Motivated by this context, the present paper means to show the potentiality and the practical applicability to clustering methods of the  functional data approach that avoids the need for arbitrary and often controversial feature extraction to find  out  homogeneous groups of DRCs, which likely pertain to spot welds sharing  common mechanical and metallurgical properties.
We intend is to provide an essential hands-on overview of  the most promising functional clustering methods, and  to apply the latter to the DRCs collected from the RSW process at hand, even if they could go far beyond the specific application hereby investigated. 
The methods analyzed are demonstrated to possibly support practitioners along the identification of  the mapping relationship between process parameters and the final quality of RSW joints as well as, more specifically, along the priority assignment  for off-line testing of welded spots and  the welding tool wear analysis.
The analysis code, that  has  been developed  through the software environment $\textsf{R}$, and the DRC data set are made openly available online at \url{https://github.com/unina-sfere/funclustRSW/}.
\end{abstract}

\noindent{\bf Key Words:} functional data analysis, functional clustering, resistance spot welding, dynamic resistance curve, industry 4.0.

\section{Introduction}
\label{sec:int}

Resistance Spot Welding (RSW) is the most common technique employed in joining metal sheets during body-in-white  assembly of automobiles\cite{osayande,zhou2014study}, mainly because of its adaptability for mass production\cite{martin}. 
Typical car body contains about 5000 spot welds joining metal sheets of different materials and thicknesses\cite{zhao2006research}. The quality of many critical spots \cite{el-banna} is routinely controlled in order to guarantee the structural integrity and solidity of  welded assemblies per vehicle \cite{martin}. 
Quality assessment  is typically based on tests performed at the end of the RSW process (off-line) on  finished sub-assemblies through direct or indirect evaluation of weld-joint key characteristics\cite{raoelison}. 
%Off-line tests are broadly grouped into destructive and non-destructive methods. Destructive tests typically allow for a direct measurement of the key spot %weld characteristics and are regarded as the most reliable and costly methods. 
%Because of their destructive nature, they can be performed only on a very small part of the production, e.g., in the tuning phase of the process to certify %that the chosen parameters guarantee the required mechanical and metallurgical properties. 
%For this reason, non-destructive techniques are undoubtedly preferred, even if they allows only for indirect evaluation of the key characteristics under %study.
%Ultrasonic tests, which pertain to the non-destructive techniques,
%are becoming more and more important in spot welding inspection\cite{roye2003ultrasonic}. However, they result still expensive for wide scale deployment on %each critical spot, and their reliability may be affected by the correct manual positioning of the probe \cite{buckley2009improvements}. Moreover, the %interpretation of oscillograms, as the main output of ultrasonic tests, is sometimes cumbersome and may depend on human operator experience and efficiency %\cite{martin}. 
Off-line testing is, however, costly and lengthy and thus  unfeasible on the full production, especially on large scale. 
%and is applied only on randomly sampled spot welds among the most critical ones on the work pieces.

In the  modern automotive industry 4.0 framework, automatic acquisition systems allow  to routinely control welders during running operations (on-line) through the continuous  record of a large volume of process parameters. In particular, the so-called \textit{dynamic resistance curve} (DRC) is the most important process parameter acquired on-line\cite{zhou} and is popularly recognized as the full technological signature of the metallurgical development of a spot weld \cite{dickinson}.

In this scenario, a paramount issue constantly faced by practitioners is the identification  of homogeneous groups (clusters) of spot welds based on DRC observations, in terms of mechanical and metallurgical properties. 
The identification of clusters with a convenient interpretation is useful for exploring the mapping relationship between process parameters  and the 
final quality of the RSW joints produced, and, in general,  for supporting the experience-based learning of any technological process.  In this regard, the most common practice in industry  is  to analyze one or few scalar features extracted from the   acquired DRC, even though feature extraction is known to be often difficult, arbitrary and risky of collapsing useful information.

On the contrary, in this paper,   each  DRC observation is suitably modelled as  a  function defined on the time domain, i.e., as \textit{functional datum}. Functional data analysis (FDA) \cite{ramsay2005functional, horvath2012inference, ferraty2006nonparametric, kokoszka2017introduction} is the set of methods that  consider functional data as its founding elements.
Clustering  functional data is usually a difficult task, because of the intrinsic infinite dimensionality of the problem.
A thorough overview of functional clustering methods can be found in Ramsay and Silverman\cite{ramsay2005functional} and  Ferraty  and Vieu\cite{ferraty2006nonparametric}. Then, it  is worth mentioning  Cuesta-Albertos and Fraiman\cite{cuesta2007impartial} who proposed  a pure functional version of the k-means algorithm, which is very popular in the multivariate setting\cite{everitt2011cluster}, as an alternative to the method of Abraham et al.\cite{abraham2003unsupervised}, who instead applied the k-means algorithm  to the coefficients obtained  by  projecting  the original profiles onto a lower-dimensional subspace spanned by B-spline basis functions. 
Another version of k-means algorithm is that of Chiou and Li\cite{chiou2007functional}, which relies  on a particular distance  between truncations at a given order of the functional principal components expansion\cite{ramsay2005functional,hall2006properties}. This version can be broadly regarded as an instance of the method proposed by  Bouveyron and Jacques\cite{bouveyron2011model}, who modelled    the functional principal components through Gaussian mixture. Some parsimony constraints on the variance parameters are also considered to define a family of
parsimonious sub-models.
A similar methods, which is based on a functional principal components expansion of the functional observations, was proposed by Jacques and Preda\cite{jacques2013funclust}.
The work of  James and Sugar\cite{james2003clustering} is recognized as  the first example of model-based procedure for functional data clustering, as well as the method proposed by Giocofci et al.\cite{giacofci2013wavelet}, which, in particular, relies on the wavelet decomposition of the functional observations, and is particularly appropriate for peak-like data, as opposed to methods based on splines.
More recently, Delaigle et al.\cite{delaigle2019clustering} proposed a functional k-means algorithm able to cluster observations asymptotically perfectly.
 A sparse functional clustering procedure, that is clustering functional data while jointly selecting the most relevant features, was developed by Floriello and Vitelli\cite{floriello2017sparse} and, in particular, by Vitelli\cite{vitelli2019novel} who  accounted for  possible curve misalignments.
For the sake of completeness, Bayesian approaches have appeared as well \cite{ray2006functional,rodriguez2009bayesian,rigon2019enriched} in the literature, even if they are beyond the scope of this paper.

After providing Section \ref{sec_techno} with  the technological background and the description of  the functional DRC data set collected from the RSW process that has motivated this research,  we give in Section \ref{sec_approaches} a deeper hands-on illustration of   the most promising functional clustering  methods to be applied to the DRC data set at hand.
In Section \ref{sec_results}, we discuss and interpret from technological viewpoint the main  results obtained, even if the proposed approach could go far beyond the specific application hereby investigated.
We conclude by Section \ref{sec_conclusions} 
with a discussion of issues highlighted by this
data set and a broader perspective of the potentiality of the proposed methods. 
Technical details for each of the clustering methods implemented in this paper are presented in the Appendix.

The DRC data set and the $\textsf{R}$\cite{Rcoreteam2020} code are 
%provided as supplementary material and are 
made openly available online 
%at \url{https://github.com/unina-sfere/funclustRSW} 
\cite{githubref} to allow the reader to possibly investigate other approaches with this data set and to encourage the fruitful spread of functional data clustering methods among practitioners in industry.

\section{Technological Background and Data Structure} 
\label{sec_techno}
The considered RSW process\cite{zhang2011resistance} refers to an autogenous welding process in which two overlapping  steel galvanized sheets are joint together, without the use of any filler material, at discrete spots. Joints are performed by applying pressure to the weld area from two opposite sides by means of two copper electrodes.
Voltage applied to the electrodes generates a current flowing between them through the material.
The electrical current flow because the resistance offered by metals causes a large heat generation (Joule effect) that increases the metal temperature at the faying surfaces of the work pieces up to the melting point.
Finally, due to the mechanical pressure of the electrodes, the molten metal of the metal sheets jointed cools and solidifies forming the so-called weld \textit{nugget}\cite{raoelison, manufacturers2003resistance}.

\begin{figure}
\centering
\includegraphics[width=.5\textwidth]{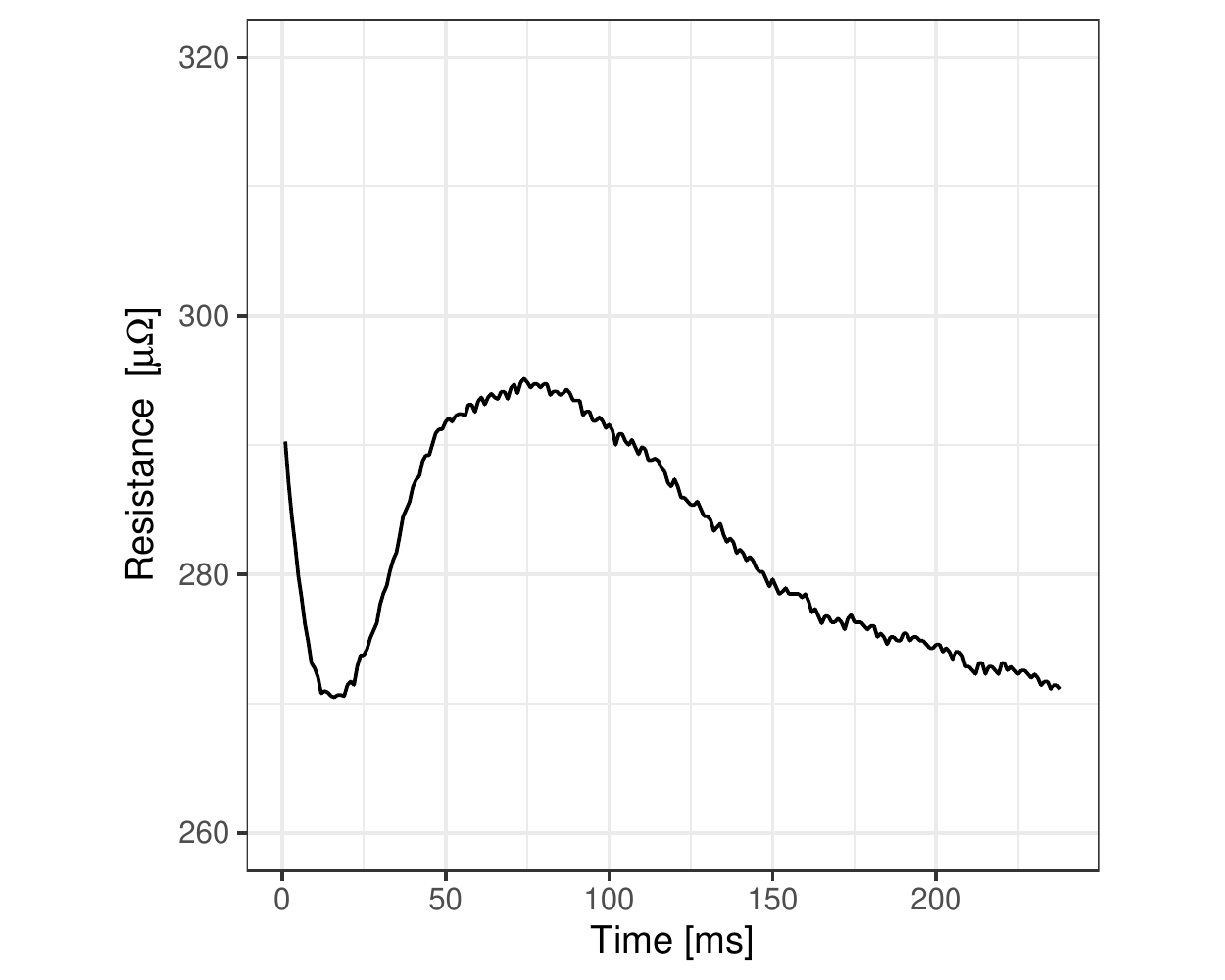}
\caption{Typical DRC behaviour.}
\label{fig_DRC_ref}
\end{figure}

The typical shape of a DRC acquired during this process is displayed in Figure \ref{fig_DRC_ref} for illustrative purposes.
In the light of Dickinson et al.\cite{dickinson}, it mainly depends on physical changes induced in the material by the ongoing welding process and can be roughly  outlined into five stages, as well depicted by Adams et al.\cite{adams2017correlating}. 
For the sake of conciseness, these stages can be  summarized as influenced by two main concurrent effects due to (\textit{a}) the metal electrical resistivity and (\textit{b}) the contact area among the metal sheets to joint. 
These effects develop during the RSW process by means of the heat  produced by the current flow and the clamping pressure generated by copper electrodes. 
In particular, DRC values are proportional to (\textit{a}), which increases with material temperature.
%Then, the temperature suddenly rises at the melting point, that ideally corresponds to the inflection point located between the local minimum and maximum of the reference DRC of Figure \ref{fig_DRC_ref}. 

On the contrary, DRC values are decreasing with \textit{b}. That, in turn, is increasing with two main factors: (\textit{b.1}) the deformation, due to the clamping force, of the surface asperities, that are softened by the high temperatures; and (\textit{b.2}) the melting of the metal, that guarantees the sheet continuity by occupying the interstices between the work pieces to weld. 
So stated, the typical DRC behaviour (Figure \ref{fig_DRC_ref})  can be interpreted by the turnover of the effects due to (\textit{a}) and (\textit{b}). Specifically, DRC decreases at first because of the effect due to  (\textit{b.1}), which dominates effect due to (\textit{a}) until the local minimum; then, conversely, DRC increases because the effect due to (\textit{a}) dominates effect due to (\textit{b}) until the local maximum, which represents the beginning of the nugget formation. Finally, the DRC decreases slowly to the end of the RSW process, because the effect due to factor (\textit{b.2}) dominates that due to (\textit{a}) to a lesser extent.
In a nutshell, DRC behaviour can be roughly outlined by one local minimum point, one local maximum point, and the resistance value at the end of the welding process.

\begin{figure}
\centering
\includegraphics[width = 0.5 \textwidth]{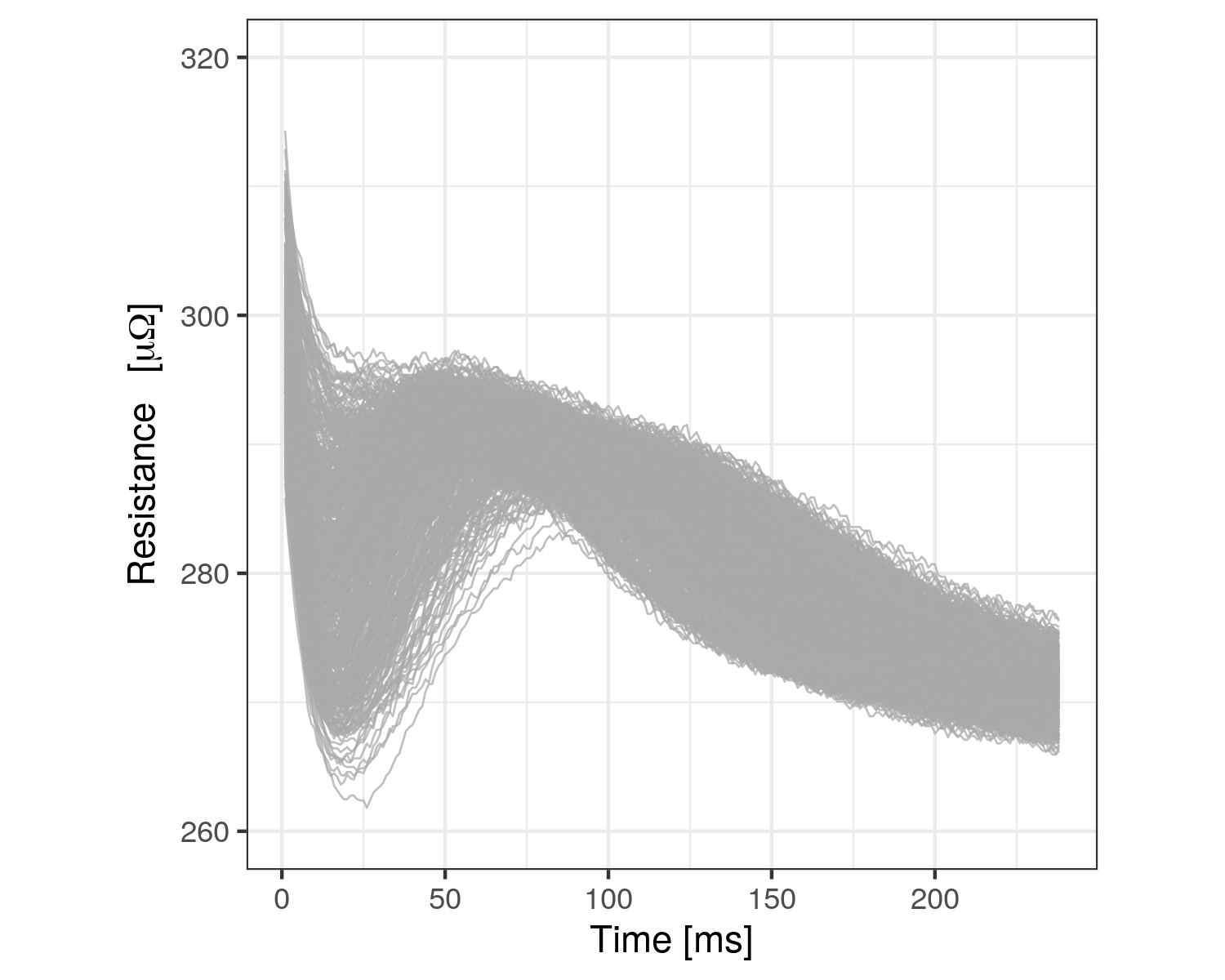}
\caption{Raw data plot of spot welding DRCs.}
\label{fig_profiles}
\end{figure}

The data set for the problem at hand consists of 538 DRCs that are plotted in Figure \ref{fig_profiles} and pertains to spot welds of the same type collected during RSW lab tests at Centro Ricerche Fiat (CRF). 
The latter have been carried out on coupons of two sheets having thickness equal to 0.7 mm and 1.3 mm and made of \textit{FE220BH} and \textit{FE600DP} galvanised steels, respectively.
The energy was supplied in a single pulse of current. The weld time period is 237 ms.
Strictly speaking, the values of electrical resistance used to obtain each DRC observation are not direct measurements, but obtained, according to the first Ohm's law\cite{ohm1827galvanische}, as the ratio between the voltage at electrode tips and the current intensity measurements. 
For each DRC observation, these have been collected at a regular grid of 238 points equally spaced by 1 ms.
In particular, the electrode tip voltage has been measured using dressed copper wires attached to the electrodes.  
Whereas, the current intensity has been measured by means of an air-core toroid in the primary of the welder transformer.
Copper wires are checked up to ensure their integrity at the beginning of every welding cycle. 
Electrical resistance of the metal sheets is assumed much larger than that of the copper electrodes
That is, the copper electrode resistance does not practically affect the measurement of metal sheet resistance.

Note that the raw data plots for the 538 DRCs at hand yield shape and features coherent with those discussed with reference to Figure \ref{fig_DRC_ref}, but show non-negligible variability that motivates the goal of the present paper in supporting practitioners to build homogeneous groups of DRCs. 
The intent is to identify through the latter spot welds having similar mechanical and metallurgical properties, and groups themselves that stand apart from one another.
In particular, clustering methods, and even more their functional version, will be of great value in this regard with the ultimate goal of guiding practitioners along the priority assignment for off-line testing of welded spots and the welding tool wear analysis.

\section{Functional Data Clustering Approaches for Dynamic Resistance Curves}
\label{sec_approaches}
Usually, functional data consist of independent realizations $X_1,\dots,X_n$ of a functional random variable $X$ with values in an infinite dimensional space, which is typically taken to be $L^2\left(\mathcal{T}\right)$, the separable Hilbert space of square integrable functions defined on the compact domain $\mathcal{T}$.
In most applications, $\mathcal{T}\subset\mathbb{R}$ and represents time,  however, multidimensional   domains could be considered as well.
Typically, $X_1,\dots,X_n$ are not entirely available  but are observed through a finite set of observation points. 
This means, only discrete observations $\lbrace X_{ij} \rbrace$ of functional observations $ \lbrace X_i \rbrace_{i=1,\dots,n}$ at time points $\lbrace t_{ij}, j=1\dots,m_i\rbrace$ are available, being $m_i$  the number of discrete points available for the $i$-th observation.
 The aim of the clustering analysis is to define $M$ partitions, i.e.,  clusters,  of the data $X_1,\dots,X_n$ such that  observations in different clusters are
as dissimilar as possible and that  observations within the same cluster are as similar as
possible. 
In the rest of this section, we describe the most promising approaches for functional clustering, which can be grouped in raw-data clustering, filtering methods, adaptive methods, and distance-based methods.

\subsection{Raw-data clustering}
\label{sec_raw}
The raw-data clustering approach consists in the clustering of discretized version $\lbrace X_{ij}\rbrace$ of the functional observations $\lbrace X_i\rbrace$   by means of classical multivariate methods.
This simple approach  does not need the reconstruction of the functional data and relies on well-established multivariate algorithms.

One of the most popular clustering algorithm is k-means. In the functional setting, k-means aims to partition the observations into $M$ clusters $C^*_1,\dots,C^*_M$ such that the within-cluster sum of squares is minimized, that is 
\begin{equation}
    \lbrace C^*_1,\dots,C^*_M\rbrace=\argmin_{C_1,\dots,C_M}\sum_{m=1}^{M}\sum_{\bm{X}_i\in C_m}\left(\bm{X}_i-\bm{\mu}_m\right)^T\left(\bm{X}_i-\bm{\mu}_m\right),
\end{equation}
where $C_1,\dots,C_M$ are all the possible observation partitions in M groups, $\bm{X}_i=\left(X_{i1},\dots,X_{im_i}\right)^T$, and $\bm{\mu}_m$ is the mean vector of the observations in $C_m$.

Hierarchical clustering\cite{everitt2011cluster} produces a representation in which clusters at each level of the hierarchy is formed by all and only the  clusters of the  lower levels.
Strategies for hierarchical clustering are mainly divided into two approaches: agglomerative (bottom-up) and divisive (top-down). 
The former starts at the bottom (i.e., each observation in one cluster) and at each level recursively merges a selected pair of
clusters into a single cluster. The latter starts at the top (i.e., all observations in one cluster) and at each level recursively splits one of the existing clusters at that level into two new clusters. Different versions of agglomerative methods arise from the choice of the intergroup dissimilarity metric, e.g., single linkage, complete linkage, average linkage. Ward (1963)\cite{ward1963hierarchical} considered hierarchical clustering procedures based on minimizing the loss of information from joining two groups.
Finally, model-based clustering assumes that the data in each  cluster is generated from a given probabilistic distribution and the combined data stems from a convex combination of these distributions.
In all the aforementioned methods, the number of clusters $M$ has to be determined.
For k-means and hierarchical clustering, this can be done based on many indices\cite{charrad2012nbclust}, e.g.,  the silhouette width\cite{rousseeuw1987silhouettes}, the gap statistic\cite{tibshirani2001estimating}, the Dunn index\cite{dunn1973fuzzy}; whereas, for model-based clustering,  information criteria, such as the Akaike information criterion (AIC), Bayesian information criterion (BIC) as well as  integrated completed likelihood (ICL) could be  used.

Analysis of raw data through classical multivariate techniques has several problems, because of the high number of  evaluation points and the strong correlation.
This may especially affect the model-based approach, that assumes non-singular  covariance matrix.
Moreover, raw-data clustering approach has the drawback of not taking into account the functional nature of the data and is not suited for curves observed at different evaluation points.
While this suggests to use approaches specifically designed for functional data, for comparison purposes we still propose the raw data approach to the clustering.
Details on multivariate clustering methods can be found in Everitt et al. (2011)\cite{everitt2011cluster}, Hastie et al. (2009)\cite{hastie2009elements} and Johnson et al. (2002)\cite{johnson2002applied}.

\subsection{Filtering methods}
\label{sec_fil}
Filtering methods rely on the reconstruction of the functional observations $ \lbrace X_i \rbrace$ from the discrete points $\lbrace X_{ij} \rbrace$. 
The most common approach\cite{ramsay2005functional} is to assume that the functional observations are embedded in a finite dimensional functional space spanned by a finite set of basis functions.
In particular, each $X_i$ can be written as 
\begin{equation}
\label{eq_1}
    X_i\left(t\right)=\sum_{k=1}^{K}c_{ik}\phi_{k}\left(t\right)=\bm{c}_{i}^T\bm{\phi}\left(t\right)\quad t\in \mathcal{T}\quad i=1,\dots,n,
\end{equation}
where $\bm{\phi}=\left( \phi_1,\dots,\phi_K \right)^T$ is the vector of  basis functions that span the $K$-dimensional subset of $L^2\left(\mathcal{T}\right)$, and $\bm{c}_{i}=\left(c_{i1},\dots,c_{iK}\right)^T $ is the $K$-dimensional  coefficient vector.
The basis functions $\lbrace \phi_j \rbrace$ can be either pre-specified, e.g. B-spline\cite{de1978practical}, Fourier\cite{ramsay2005functional}, and wavelet\cite{walnut2013introduction}, or data-adaptive, e.g. obtained using functional principal component analysis (FPCA)\cite{hall2006properties}.

In case of pre-specified basis functions, if the $\lbrace X_{ij} \rbrace$ are observed with measurement error, then the  coefficient vector $\bm{c}_i$ is usually estimated as  $\hat{\bm{c}}_{i}$  via penalized least-squares, even though standard least-squares could be used as well\cite{ramsay2005functional}, that is 
\begin{equation}
   \hat{\bm{c}}_{i}=\argmin_{\bm{c}_i\in\mathbb{R}^K} \sum_{j=1}^{m_i}\left( X_{ij}-\bm{c}_i^T\bm{\phi}\left(t_{ij}\right)\right)^2+\lambda\int_{\mathcal{T}}[D^2 \bm{c}_i^T\bm{\phi}\left(t\right)]^2dt,
\end{equation}
where $D^2$ is the second order differential operator and $\lambda>0$ is a smoothing parameter. It  measures the trade-off between fit to the data, as determined by the residual sum of squares
in the first term, and smoothness of $X_i$, as quantified by  the second term.
Then, the reconstructed functional observation is
\begin{equation}
    \hat{X}^{PS}_i\left(t\right)=\hat{\bm{c}}_{i}^T\bm{\phi}\left(t\right)\quad t\in \mathcal{T}\quad i=1,\dots,n.
\end{equation}
The choice of the smoothing parameter $\lambda$ is based on the well-known  trade-off between variance and bias. In particular, it is usually  performed by picking the $\lambda$ corresponding to the minimum value assumed by the generalized cross-validation  criterion. This criterion takes into account the degrees of freedom of the estimated curve that vary according to $\lambda$ \cite{ramsay2005functional}.
Moreover, the choice of $K$ in Equation \eqref{eq_1} is not crucial\cite{cardot2003spline}, until it is  sufficiently large  to capture  local behaviours of  functional data.

The FPCA provides a  data-adaptive basis to obtain the functional data as in Equation \eqref{eq_1}.
In particular, the functional observations are reconstructed, for $i=1,\dots,n$, as
\begin{equation}
\label{eq_pca}
    \hat{X}^{DA}_i\left(t\right)=\sum_{l=1}^{L} \xi_{il}\psi_l\left(t\right)=\bm{\xi_{i}}^T\bm{\psi}\left(t\right)\quad t\in\mathcal{T}\quad i=1,\dots,n,
\end{equation}
where $\bm{\xi_{i}}=\left(\xi_{i1},\dots,\xi_{iL}\right)^T$ is the vector of  principal component scores or simply scores defined as $
      \xi_{il}=\int_{\mathcal{T}}\psi_l\left(t\right)X_i\left(t\right)dt$, and 
$\bm{\psi}=\left(\psi_{1},\dots,\psi_{L}\right)^T$ is the vector whose elements are weight functions referred to as principal components.
Principal components are defined by an iterative algorithm which at each step finds the weight function that maximizes the mean square of the scores, or their sample variance, that is 
\begin{equation}
    \psi_l=\argmax_{\psi}  \sum_{i=1}^{n}\xi_{il}^2=\sum_{i=1}^{n}\left(\int_{\mathcal{T}}\psi\left(t\right)X_i\left(t\right)dt\right)^2\quad l=1\dots,L,
\end{equation}
under the constraints: $\int_{\mathcal{T}}\psi_l\left(t\right)^2dt=1$  and $\int_{\mathcal{T}}\psi_i\left(t\right)\psi_j\left(t\right)dt=0$, for $i\neq j$.
The choice of the number $L$ in Equation \eqref{eq_pca} of retained components  depends on several necessities. Generally, the retained principal components are chosen such that they explain at least a given percentage of the total variability. However, more sophisticated methods could be used as well\cite{jolliffe2011principal}.
In practice,  reconstruction of functional observations allows one  to reduce the dimensionality of the data by summarizing each curve through a finite set of parameters, that is $\lbrace \hat{\bm{c}}_i \rbrace$ or $\lbrace \bm{\xi}_i \rbrace$ depending on whether basis functions  used are pre-specified or data-adaptive.
Then, the finite set of parameters are clusterized  by means of standard multivariate clustering techniques, such as k-means, hierarchical clustering or model-based clustering. As for the raw-data clustering  methods, several indices  could be used\cite{charrad2012nbclust} to choose the number $M$ of clusters.

\subsection{Adaptive methods}
\label{sec_adme}
The present set of methods relies on a finite dimensional representation of the functional data through basis functions (similarly to the filtering approaches) where the basis expansion coefficients are  treated as random variables with cluster-specific probability distributions.
This differs  from the filtering methods, where the basis expansion coefficients are considered as parameters.
One of the first example of adaptive method  was in James and Sugar\cite{james2003clustering}, referred to as \textit{fclust}.
Similarly to the filtering approaches,  if the functional observation $X_i$ belongs to the $m$-th cluster among the $M$ clusters, it is modeled through basis functions as
\begin{equation}
     X_i\left(t\right)=\bm{\eta}_{im}^T\bm{\phi}\left(t\right)\quad t\in \mathcal{T}\quad i=1,\dots,n,
\end{equation}
where $\bm{\phi}=\left( \phi_1,\dots,\phi_K \right)^T$  are natural cubic splines, and $\bm{\eta}_{ik}$ is a vector of spline normal random coefficients defined as 
\begin{equation}
 \bm{\eta}_{im}=\bm{\mu}_m+\bm{\gamma}_i,
\end{equation}
with $\bm{\mu}_m$  the coefficient vector of the $m$-th cluster mean, and $\bm{\gamma}_i\sim N\left(0,\bm{\Gamma}\right)$ the  subject-specific random effects for the $i$-th curve.
 Then, the vector of discretized values $\bm{X}_i=\left(X_{i1},\dots,X_{im_i}\right)^T$ is modelled as
\begin{equation}
    \bm{X}_i=\bm{S}_i\left(\bm{\mu}_m+\bm{\gamma}_i\right)+\bm{\varepsilon}_i,
\end{equation}
where $\bm{S}_i=\left(\bm{\phi}\left(t_{i1}\right),\dots,\bm{\phi}\left(t_{im_i}\right)\right)^T$ is the realization matrix of the vector $\bm{\phi}$, and $\bm{\varepsilon}_i\sim N\left(0,\bm{R}\right)$ is the measurement error random vector. The covariance matrix $\bm{R}$ is usually set equal $\sigma^2\bm{I}_{m_i}$, where $\bm{I}_{m_i}$ is the size ${m_i}$ identity matrix.
The unknown parameters $\bm{\mu}_m$,  $m=1,\dots,M$, $\bm{\Gamma}$ and $\sigma$ are estimated by maximizing the mixture likelihood in Equation \eqref{eq_mixli}, where the cluster membership vector is modeled as a multinomial random variable with parameters $\left(\pi_1,\dots,\pi_M\right)$, with $\pi_m$ the probability of an observation to belong to the $m$-th cluster. Thus, the mixture likelihood is defined as
\begin{equation}
\label{eq_mixli}
    L\left(\bm{\mu}_1,\dots,\bm{\mu}_M,\bm{\Gamma},\sigma,\pi_1,\dots,\pi_M\right)=\prod_{i=1}^{N}\sum_{m=1}^{M}\pi_mf_m\left(\bm{X}_i\right),
\end{equation}
where $f_m\left(\bm{X}_i\right)$ is the conditional density  function of $\bm{X}_i$ belonging to the $m$-th cluster, that is $\bm{X}_i|m\sim N\left(\bm{S}_i\bm{\mu}_m,\bm{\Sigma}_i\right)$, with $\bm{\Sigma}_i=\sigma^2\bm{I}_{m_i}+\bm{S}_i\bm{\Gamma}\bm{S}_i^T$.
The maximization is often carried out by means of the expected maximization (EM) algorithm.
Once the unknown parameters have been estimated, each curve $X_i$ is assigned to the cluster whose estimated posterior probability of cluster membership $\pi_{m|i}=\hat{f}_m\left(\bm{X}_i\right)\hat{\pi}_m/\sum_{j=1}^{M}\hat{f}_j\left(\bm{X}_i\right)\hat{\pi}_j$ is maximum.
Moreover, the cluster mean coefficients $\bm{\mu}_m$ could be further optimally parameterized to produce useful low-dimensional representations of the curves\cite{james2003clustering}.
Information criteria, such as AIC and BIC, are used to select the number $M$ of clusters  and the basis dimension $K$\cite{james2003clustering}.

The use of spline basis has two main drawbacks: ($i$) they are inappropriate when dealing with functions that show peaks and irregularities, ($ii$) they require heavy computational efforts and so
 are not suitable to represent high dimensional data.
 For these reasons, Giocofci et al.\cite{giacofci2013wavelet} proposed an adaptive method based on the wavelet decomposition of the curves, referred to as \textit{waveclust}.
 Similarly to James and Sugar\cite{james2003clustering},   the functional observation $X_i$ belonging to the $m$-th cluster is modeled as 
\begin{equation}
\label{eq_modwav}
    X_i\left(t\right)=\mu_m\left(t\right)+U_i\left(t\right)\quad t\in \mathcal{T}\quad i=1,\dots,n,
\end{equation}
where $\mu_m$ is the principal functional fixed effect  that characterizes the $m$-th cluster mean and $U_i$ is a subject-specific random deviation from   $\mu_m$.
By applying discrete  wavelet transform to model in Equation \eqref{eq_modwav}, contaminated with  an additional  measurement error function  $E_i\left(t\right)$, $t\in \mathcal{T}$, the model reduces to a linear mixed-effect one. That is,
\begin{equation}
\label{eq_projwave}
\left(\bm{c}_i^T,\bm{d}_i^T\right)^T=\left(\bm{\alpha}_m^T,\bm{\beta}_m^T\right)^T+\left(\bm{\nu}_i^T,\bm{\theta}_i^T\right)^T+\left(\bm{\varepsilon}_{\bm{c}_i}^T,\bm{\varepsilon}_{\bm{d}_i}^T\right)^T,
\end{equation}
where $\left(\bm{\alpha}_m^T,\bm{\beta}_m^T\right)^T$, $\left(\bm{\nu}_i^T,\bm{\theta}_i^T\right)^T$, $\left(\bm{\varepsilon}_{\bm{c}_i}^T,\bm{\varepsilon}_{\bm{d}_i}^T\right)^T$, and $\left(\bm{c}_i^T,\bm{d}_i^T\right)^T$ are the vectors of scaling and wavelet coefficients of $\mu_m$, $U_i$, $E_i$ and $X_i+E_i$, respectively; $\bm{\alpha}_m$ and $\bm{\beta}_m$ are non-random parameters, whereas $\left(\bm{\nu}_i^T,\bm{\theta}_i^T\right)^T$ and $\left(\bm{\varepsilon}_{\bm{c}_i}^T,\bm{\varepsilon}_{\bm{d}_i}^T\right)^T$ are  normal random vectors with zero mean and covariance matrices  $\bm{G}$ and $\sigma_\varepsilon\bm{I}_{m_i}$, respectively.  Once projected in the wavelet domain, the clustering model \eqref{eq_projwave}  resumes to a standard one with additional  random effects whose variance is of particular form. Thus, parameters are estimated by maximizing the  likelihood function typically  using the
 EM algorithm. Final assignment of each curve to a cluster  is performed by maximizing the   posterior probability of clustering membership. The number of clusters are chosen through BIC or ICL\cite{giacofci2013wavelet}.
 
 The last presented adaptive method was proposed by  Bouveyron and Jacques\cite{bouveyron2011model}, and referred to as \textit{funHDDC}. They consider, as  James and Sugar\cite{james2003clustering}, that if $X_i$ belongs to a given cluster $m$, it admits the following basis expansion
 \begin{equation}
     X_i\left(t\right)=\bm{\gamma}_{im}^T\bm{\Psi}\left(t\right)\quad t\in \mathcal{T}\quad i=1,\dots,n,
 \end{equation}
where $\bm{\Psi}=\left( \Psi_1,\dots,\Psi_K \right)^T$  is a given vector of basis functions, and $\bm{\gamma}_{im}$ is a $k$-dimensional random vector.
All the functions $X_i$ in a given cluster $m$ are assumed to be adequately described in a low-dimensional functional latent subspace with dimension $d_m<K$ spanned by a group-specific basis function $\lbrace\varphi_{mj}\rbrace$.
Then, for a given $X_i$ in the cluster $m$, the  random latent expansion coefficients $\bm{\lambda}_i=\left(\lambda_{i1},\dots,\lambda_{id_m}\right)^T$  in the group-specific basis function $\lbrace\varphi_{mj}\rbrace$ are linked to $\bm{\gamma}_{im}$ as
\begin{equation}
    \bm{\gamma}_{im}=\bm{U}_m\bm{\lambda}_i+\bm{\varepsilon}_i,
\end{equation}
where $\bm{U}_m$ is the $K\times d_m$ matrix composed by the first $d_m$ columns of the orthogonal $K\times K$ matrix $\bm{Q}_m$, whose entries are the  coefficients that the linearly link $\lbrace \Psi_k\rbrace$ and $\lbrace\varphi_{mj}\rbrace$, and $\bm{\varepsilon_i}\in\mathbb{R}^K$ is an independent random noise term.
By assuming that $\bm{\lambda}_i\sim N\left(\bm{\mu}_m,\bm{S}_m\right)$, with  $\bm{S}_m=\diag\left(a_{m1},\dots,a_{md_m}\right)$, and that $\bm{\varepsilon}_i\sim N\left(0,\bm{\Xi}_m\right)$, then
\begin{equation}
\label{eq_gamma_i}
    \bm{\gamma}_{im}\sim N\left(\bm{U}_m\bm{\mu}_m,\bm{Q}_m\bm{\Delta}_m\bm{Q}_m^T\right),
\end{equation}
where the $K\times K$ matrix  $\bm{\Delta}_m=\bm{Q}_m\left(\bm{U}_m\bm{\Delta}_m\bm{U}_m^T+\bm{\Xi}_m\right)\bm{Q}_m^T$ and the noise covariance matrix $\bm{\Xi}_m$ is chosen such that  $\bm{\Delta}_m=\diag\left(a_{m1},\dots,a_{md_m},b_m,\dots,b_m\right)$.
Let us assume  the cluster membership vector is modeled as a multinomial random variable with parameters $\left(\pi_1,\dots,\pi_M\right)$, with $\pi_m$ the probability of an observation to belong to the $m$-th cluster. Then, the mixture likelihood is defined as
\begin{equation}
    L\left(\bm{U}_1,\dots,\bm{U}_M,\bm{\mu}_1,\dots,\bm{\mu}_M,\pi_1,\dots,\pi_M,\bm{Q}_1,\dots,\bm{Q}_M,\bm{\Delta}_1,\dots,\bm{\Delta}_M\right)=\prod_{i=1}^{N}\sum_{m=1}^{M}\pi_mf_m\left(\bm{\gamma}_i\right),
\end{equation}
where $f_m\left(\bm{\gamma}_i\right)$ is the conditional density  function of $\bm{X}_i$ to belong  to the $m$-th cluster, that is $\bm{\gamma}_i|m\sim N\left(\bm{U}_m\bm{\mu}_m,\bm{Q}_m\bm{\Delta}_m\bm{Q}_m^T\right)$.  The maximization is conveniently  carried out by means of the  EM algorithm.
Moreover, it is possible to obtain parsimonious submodels of Equation \eqref{eq_gamma_i}  by constraining model parameters within or between groups.
The latent subspace dimension $d_m$ and the number of clusters $M$ are chosen through a scree-test and BIC, respectively\cite{bouveyron2011model}.

\subsection{Distance-based methods}
These methods are the functional extension of classical geometric clustering algorithm to functional data, such as k-means\cite{cuesta2007impartial} and hierarchical\cite{ferraty2006nonparametric} clustering, that basically rely on the definition of proximity or dissimilarity among observations. Therefore, the extension to functional data consists in  the introduction of an appropriate functional measure of proximity or dissimilarity.
In this respect, several authors\cite{tarpey2003clustering,ferraty2006nonparametric,cuesta2007impartial} agree upon the use  of the following measure  of proximity between the curves $X_i $ and $X_j$
\begin{equation}
\label{eq_dist}
  d_l\left(X_i,X_j\right)=\left(\int_{\mathcal{T}}\left(X_i^{\left(l\right)}-X_j^{\left(l\right)}\right)^2dt\right)^{1/2},
\end{equation}
where $X^{\left(l\right)}$ denotes the $l$-th derivative of $X$.
In this case the number of clusters could be suitably chosen through  the silhouette index\cite{rousseeuw1987silhouettes}.

\section{Results and Discussion}
\label{sec_results}
%\begin{itemize}
%    \item slide christian sui risultati
%    \item clusters e centroidi 
%    \item spiegazione dei centroidi
%    \item indirizzare il controllo off-line a ultrasuoni o la sostituzione %dell'elettrodo, carta di controllo per l'utilizzo on-line (in generale)
%\end{itemize}
In this section, we discuss on the results obtained by implementing the functional clustering methods presented in Section \ref{sec_approaches} to the DRC data set illustrated in Section \ref{sec_techno}.
For the sake of readability, implementation details are deferred to the Appendix.
The optimal number of clusters selected by each approach mentioned in Section \ref{sec_approaches} is reported in Table \ref{tab_clusters}. 
\begin{table}
\caption{Number of clusters obtained and computation time for each approach. Programs were run using a machine with an Intel Xeon 2.10 GHz processor.}
\begin{tabular}{ccc}
Method                          & Number of clusters & Computation time (min) \\ \hline
Raw data hierarchical           & 2                  & \multirow{3}{*}{3}     \\
Raw data k-means                & 3                  &                        \\
Raw data model-based            & 8                  &                        \\ \hline
Filtering B-spline hierarchical & 3                  & \multirow{3}{*}{2}     \\
Filtering B-spline k-means      & 3                  &                        \\
Filtering B-spline model-based  & 4                  &                        \\ \hdashline
Filtering FPCA hierarchical     & 3                  & \multirow{3}{*}{1}     \\
Filtering FPCA k-means          & 3                  &                        \\
Filtering FPCA model-based      & 5                  &                        \\ \hline
Adaptive fclust                 & 4                  & 621                    \\
Adaptive curvclust              & 2                  & 359                    \\
Adaptive funHDDC                & 2                  & 559                    \\ \hline
Distance-based                  & 2                  & 2                      \\ \hline
\end{tabular}
\label{tab_clusters}
\end{table}
Note that most of the methods provide similar results and identify two or three clusters. 
The only exceptions are some \textit{model-based} methods, viz. \textit{adaptive fclust}, \textit{filtering B-spline}, \textit{filtering FPCA}, and \textit{raw data}, which select from four to eight clusters.
In general, the larger the number of clusters, the harder the technological interpretation, i.e., the less straightforward the discrimination of spot welds belonging to different groups.
Inflation in the number of clusters is usually due to overfitting problems especially for model-based approaches applied to high-dimensional correlated data and complex variance structures. 
In this case, the number of parameters to be estimated can be very large and may lead to instability, no matter if the BIC criterion, that penalizes the model complexity, is used to select the optimal number of clusters.
This issue may be exacerbated by the use of model-based methods on raw data (see third row of Table \ref{tab_clusters}), which do not rely on an optimal basis representation and typically contain additional noise.
% We recommend avoiding methods not specifically appropriate for this data set.
% For example, although providing two clusters with similar results to the other methods, \textit{clurvclust} is meant to be used with wavelets, which are well suited for identifying highly discriminant local time and scale features, while this data set consists of smooth functions.
In Table \ref{tab_clusters} it is also reported the computation time required for each approach using a machine with an Intel Xeon 2.10 GHz processor.
The adaptive approaches result as the most computationally intensive, while all the others require less than 3 minutes to complete the analysis.
Even if strictly dependent on the data set at hand, this information may be crucial when dealing with complex data structures in order to pick the most appropriate method to be used when computational resources are limited.

\begin{figure} 
\centering
\includegraphics[width=\textwidth]{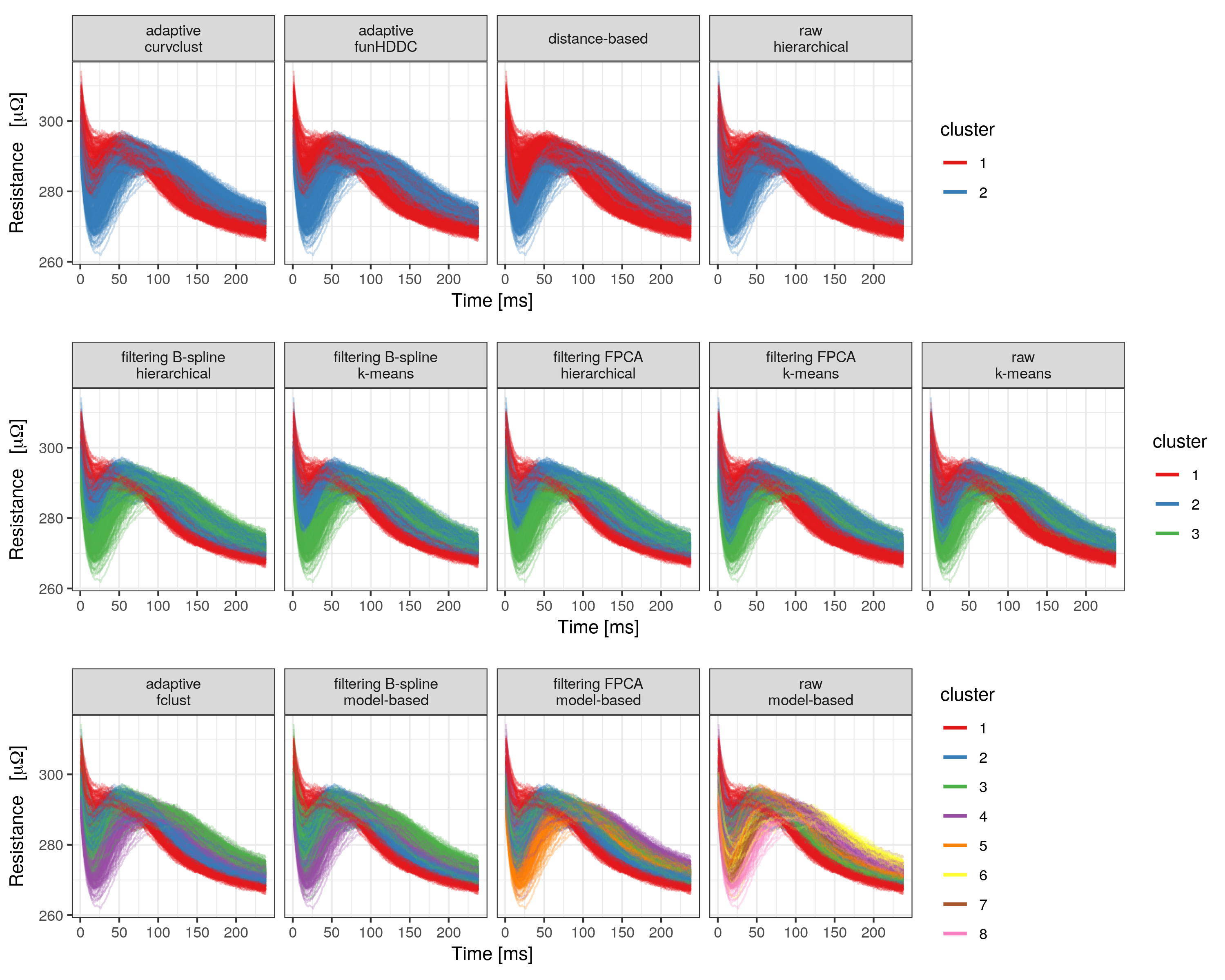} 
\caption{Plot of the functional data. Each panel correspond to one of the proposed clustering methods, curves are coloured accordingly to the corresponding cluster assignment.
Plots are arranged such that first, second and third rows of panels refer to clustering methods that divide  DRCs into two, three, and more than three clusters, respectively.}
\label{fig:col_functions}
\end{figure}

\begin{figure} 
\centering
\includegraphics[width=\textwidth]{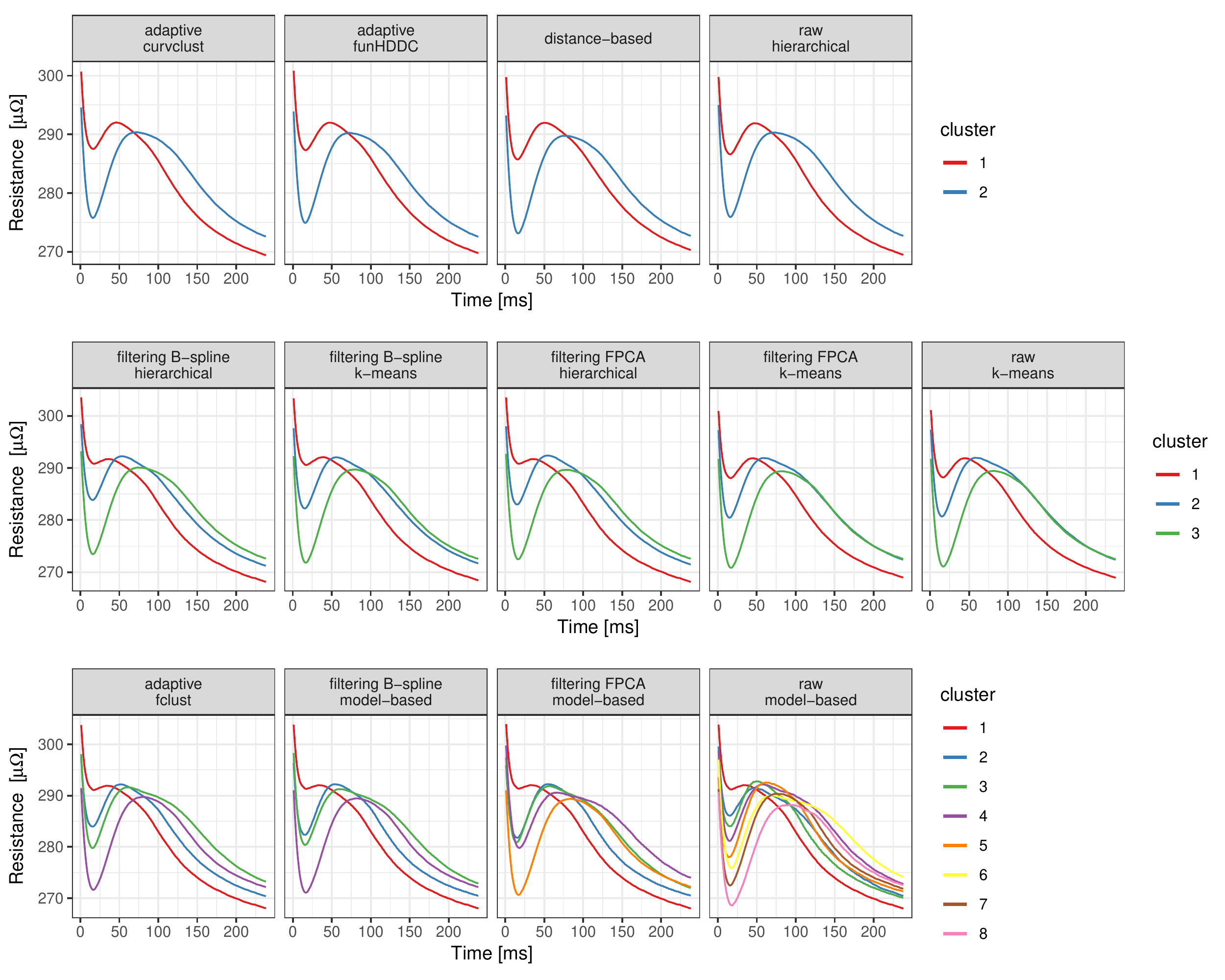} 
\caption{Plot of the cluster centroids. Each panel corresponds to one of the proposed clustering methods, curves are centroids of each cluster obtained with the corresponding method.
Plots are arranged such that first, second and third rows of panels refer to clustering methods that divide  DRCs into two, three, and more than three clusters, respectively.}
\label{fig:centroids}
\end{figure}

Figure \ref{fig:col_functions} shows the DRCs coloured according to the cluster assignment provided by each method. 
Whereas, Figure \ref{fig:centroids} depicts the centroids (i.e., mean functions) for each cluster.
Note that, in both figures, first, second and third rows of panels refer to clustering methods that select two, three and more than three clusters, respectively.


%%inizio spiegazione centroidi%%%%%%%%%%%%%%%%%%
%DRC centroids depicted in Figure \ref{fig:centroids} show important features and landmarks that allow technological interpretation consistent with the literature\cite{dickinson,adams2017correlating} and valuable  insights on the industrial problem at hand.
%In the following we provide the interpretation of the different clusters by associating each group to the size of the welding nugget obtained. 
%Clusters with a higher minimum value in the first part of the functional domain tend to have a larger welding nugget. 
With reference to those figures and coherently  with the features highlighted in Section \ref{sec_techno}, we note that DRC centroids have local minimum points with approximately the same abscissa, but different resistance values; local maximum points with approximately the same value, but different abscissa; different resistance values at the end of the functional domain.
It will be in fact convenient to facilitate the following technological interpretation and  insights into the industrial problem at hand to focus attention on  (\textit{I}) the \textit{amplitude difference} between minimum and maximum resistance values, (\textit{II}) the \textit{phase difference} between minimum and maximum abscissas, and (\textit{III}) the \textit{final resistance value}.

\begin{figure}
\centering
\includegraphics[width=.5\textwidth]{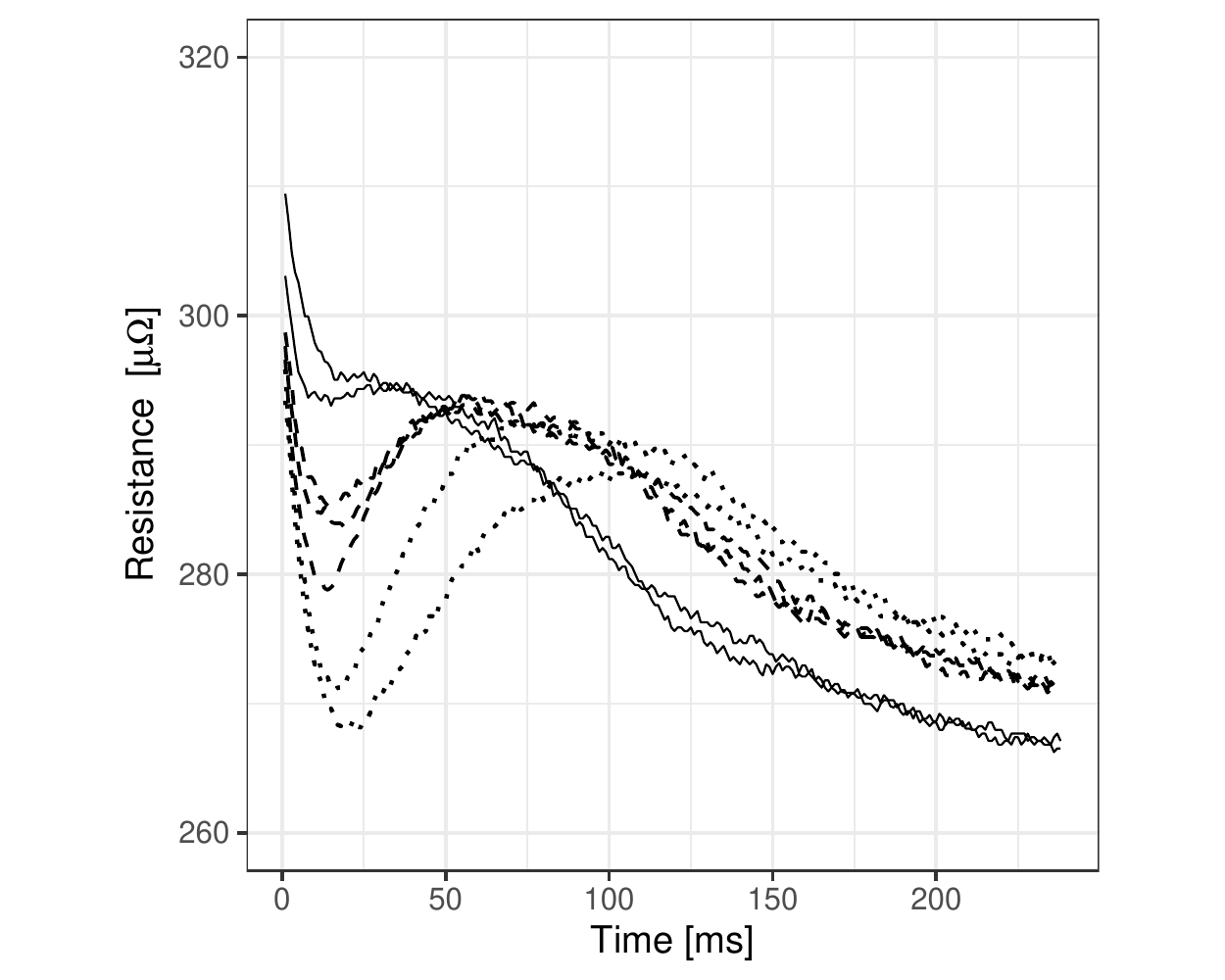}
\caption{Seven DRC observations, included in the 538 original DRCs, corresponding to spot welds for which qualitative information about the electrode wear status is available: just renewed (thin solid line), intermediate wear (dashed line), severe wear (dotted line).}
\label{fig:wear}
\end{figure}

The first row of panels of Figure \ref{fig:centroids} displays centroids associated to cluster 1 having amplitude, phase difference and final resistance value smaller than those respectively associated to cluster 2. 
The separation is clear as cluster 1 centroids show a local minimum value, in the first part of the functional domain, that is distinctly larger than that corresponding to cluster 2 centroids, and decrease more rapidly to lower values in the last part of the domain.

Whereas, in the second row of panels, the amplitude, phase difference and final value of centroids increase together from cluster 1 to cluster 3, except for  panels \textit{filtering FPCA k-means} and \textit{raw k-means} that have centroids of clusters 2 and 3 with approximately the same final value. 
That is, cluster 1 centroids show the larger local minimum value and the more rapid decrease at the end of the functional domain, whereas the other centroids tend to be more similar.

Finally, with reference to each panel of the third row of Figure \ref{fig:centroids}, centroids are shown to have  cluster number sorted in ascending order by amplitude and phase difference, only. 
That is, the final resistance values do not preserve the order  set by the phase difference, as in the first two rows of panels.
In fact, with reference to the third row of panels of Figure \ref{fig:centroids}, final resistance values of centroids obtained by \textit{adaptive fclust} and \textit{filtering B-spline model-based}, respectively depicted in the first and second panel, are sorted in ascending order with the cluster number as 1,2,4,3; whereas, those of the third panel, referring to \textit{filtering FPCA model-based} method, are ordered as 1,2,3,5,4; and those of the fourth panel (\textit{raw model-based} method) as 1,3,2,5,7,8,4,6.

As it typically happens\cite{everitt2011cluster}, also for this data set no one clustering method can be judged to be best in all circumstances. However, dealing with a small number of clusters, say two or three in this case, shall provide with a clearer interpretation of groups of functions that are well distinct and
%with the aim to provide the clustering techniques 
 more likely to lead to informative classifications.
%i punti con res finale più bassa sono migliori se a parità di altre condizioni, che qui non sono uguali a causa dell'usura. quindi nocciolo più grande è megli ma sempre entro certi limiti, per esempio lo spruzzo ha un valore più basso ma non vanno bene. 
%phase maggiore più a destra meno il nocciolo ha tempo per crescere (minore tempo tra il massimo e la fine).
% amplitude (differente superficie elettrodo) maggiore comporta uno shift maggiore sempre ...
%To illustrate technological insight triggered by the functional cluster analysis and the electrode wear information depicted in Figure \ref{fig:wear}, without loss of generality, 
Therefore, in what follows we assume selecting three clusters as the better compromise to trade off  straightforward interpretation of DRCs belonging to the same clusters and distinct characterization of each cluster.

Consistently with the technological literature\cite{dickinson, adams2017correlating}, being the minimum point abscissa practically constant and the maximum point the landmark for the start of nugget formation (see also Section \ref{sec_techno}), we can state that the smaller the phase difference, and thus the larger the time interval between the local maximum and the end of the welding process, the more the heat energy supplied for nugget growth. 
Note that the inflection point typically located between the local minimum and maximum of the DRC (see, e.g. Figure 1) ideally represents the welding melting point.
Centroids having larger phase difference shall thus characterize welding spot clusters with smaller nugget size.
Unfortunately, in the real case at hand, this ideal statement may not hold because of the natural wear process of the electrodes, which induces, as conjectured by RSW process experts, a non-negligible increase of the weld area, and thus different welding conditions for each spot weld. 

In order to explain the cluster in terms of electrode wear, in Figure \ref{fig:wear} we report seven DRCs for which it has been possible to retrieve qualitative information on the wear status of the electrodes. 
For this purpose, without loss of generality, we refer to the \textit{filtering B-spline hierarchical} method  among those selecting three clusters, and compare the corresponding panel in the second row of Figure \ref{fig:centroids} with Figure \ref{fig:wear}. 
With reference to the latter figure, we may want to analyze the two extreme wear cases (thin solid and dotted lines) and conjecture that centroid of cluster 1 of  Figure \ref{fig:centroids} shall correspond to the smaller electrode area, i.e., electrode just renewed, whereas centroid of cluster 3 to the larger one, i.e., electrode with severe wear.
Even though the technological cause is different, experts' opinion is that DRCs associated to cluster 1 and those associated to cluster 3 
%achieve the better spot weld quality since both 
correspond to spot welds with improper nugget diameter.
In particular, for spot welds that belong to DRCs in cluster 1, the root cause is attributed to the excessive clamping pressure in the welding zone.
%The larger current density triggered by the smaller electrode area may in fact result still insufficient for the improper clamping pressure\cite{dickinson}. 
The large clamping pressure is also  confirmed by the small amplitude difference in the DRC centroid of cluster 1\cite{dickinson}. 
%Moreover, note also that, coherently with the second Ohm's law, the latter centroid has larger resistance values than the other centroids of the same panel in the first part of the domain, i.e., until the local maximum, as  the lower the weld area the larger the resistance value. 

Conversely, spot welds pertaining to cluster 3 correspond to  larger electrode area, so that the clamping force generates the lower pressure in the welding zone and  cannot guarantee the proper value of current density. Indeed, despite the larger amplitude difference in the DRC centroid, which proves the smaller pressure, the nugget diameter may result undersized because of  the delay in the nugget formation.  
Therefore, it turns out that the better spot welds in terms of nugget formation achieved by DRCs pertaining to intermediate cluster(s). 
%The latter have in fact a proper amplitude difference and a small phase difference.

\begin{figure}
\centering
\includegraphics[width=.6\textwidth]{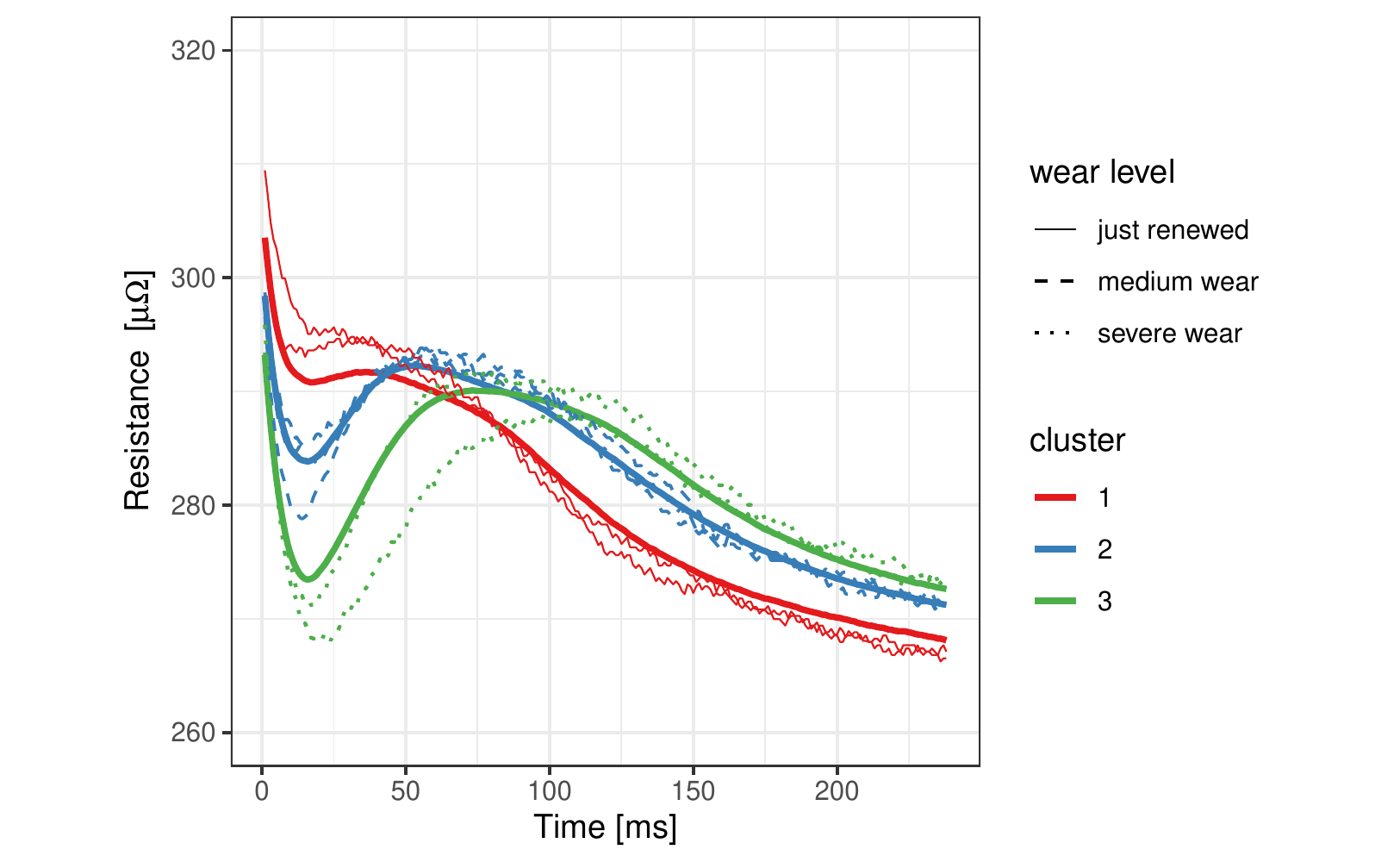}
\caption{The corresponding centroids (solid line), already displayed in the first panel of the second row of Figure \ref{fig:centroids}, are superimposed to the seven DRCs with qualitative electrode wear status: just renewed (thin solid line), intermediate wear (dashed line), severe wear (dotted line), already displayed in Figure \ref{fig:wear}, are colored by cluster number assigned through \textit{filtering B-spline hierarchical} method. }
\label{fig:wear_all2}
\end{figure}

This conjecture becomes more clear in the light of Figure \ref{fig:wear_all2} in which observations with qualitative electrode wear status, already displayed in Figure \ref{fig:wear}, are colored by cluster number (assigned through \textit{filtering B-spline hierarchical} method) and are superimposed to the corresponding centroids, already displayed in the first panel of the second row of Figure \ref{fig:centroids}. 
By Figure \ref{fig:wear_all2}, the wear status of the electrode clearly appears as a determinant factor in the clustering of DRCs. 
This result represents an important industrial finding that confirms an expert conjecture supported by functional data made available under the Industry 4.0 paradigm.
The next steps to exploit this result is to explicitly identify conditions on the wear status to signal, when the corresponding DRC is associated to a cluster that does not guarantee proper mechanical and metallurgical properties, the need e.g., for electrode break-in, renewal or substitution.
To put into action this strategy, or even more complicated maintenance programs, further technological investigation and off-line quality testing should be carried out for every DRC cluster.
The ultimate goal is to avoid the random sampling of the sub-assemblies to be tested off-line and to support more specific priority assignment.
One could in fact imagine to assign higher priority to future spot welds having DRC observation with larger (resp., smaller) distance from cluster centroids that have revealed to refer to adequate (resp., inadequate) quality.

%\newpage
%
%\begin{figure}[h!]
%\centering
%\includegraphics[width=\textwidth]{Immagini/plot_pad_wrap.pdf}
%\caption{Seven DRC observations, among the 538 in the data set, corresponding to spot welds performed by the same electrodes for which CRF has been able to provide qualitative %information about the electrode wear status: just renewed (solid line), intermediate wear (dashed line), severe wear (dotted line).}
%\label{fig:wear_all}
%\end{figure}

\section{Conclusions}
\label{sec_conclusions}
In this article, we tackled the issue of finding homogeneous groups of dynamic resistance curves (DRCs) coming from a resistance spot welding (RSW) process, which, in the modern automotive industry 4.0, is of crucial relevance to better understand the effects of the process parameters on the final weld quality.
To avoid loss of information caused by arbitrary scalar feature extraction,  DRCs have been modelled as functional data defined on the time domain, and, accordingly, 
clustering methods specifically designed for functional data have been  presented in a practical hands-on overview with the aim of facilitating their practical implementation.
To the best of the authors' knowledge, 
%This study is the first attempt to gain technological insights on  an RSW process by using all the information available in the DRCs through functional data analysis methods.
this is the first study where functional clustering methods are applied to the whole DRC functional observations
to  gain technological insights on RSW processes, 
%interpret DRCs coming from a RSW process, 
even if the framework used could go far beyond the specific application hereby investigated.

The effectiveness of the presented functional clustering methods is demonstrated by applying them to 538 DRCs acquired during RSW lab tests at Centro Ricerche Fiat (CRF). 
It turned out that the identified clusters of  DRCs  are  strictly linked with the wear status of the electrodes, that, in turn, affects the electrode contact area, clamping pressure in the welding zone and  current density, and impacts on the final quality of spot welds in terms of mechanical and metallurgical properties.  
Indeed, in accordance with the experts, we agree the better spot welds shall correspond to DRCs belonging intermediate clusters having  proper amplitude difference and  small phase difference.

A broader perspective of the results is given in supporting practitioners in the priority assignment  for off-line testing of welded spots and in the electrode wear analysis.
Functional  clustering analysis could be in fact imagined to be embedded in a wider on-line statistical quality control framework for RSW processes, which is able to properly exploit the  properties of the clusters identified. 
Finally, the relationship between the electrode wear and the final  quality of spot welds, which has been discovered by the proposed functional clustering analysis, could be now further investigated through the specific definition of opportune quantitative variables in the direction of routinely tracing wear status.

\section*{Acknowledgements}
The present work was developed within the activities of the project ARS01\_00861 ``Integrated collaborative systems for smart factory - ICOSAF'' and financially supported by MIUR (Ministero dell'Istruzione, dell'Università e della Ricerca).
The authors are extremely grateful to  CRF  (\texttt{www.crf.it}) Engineers Alessandro Zanella, Gianmarco Genchi, and Mariano Quagliano for their technological insights in the interpretation of the results.

\bibliographystyle{qrei}
\bibliography{References}

\section* {Appendix A: Implementation Details}

In the following paragraphs, we provide with further details for each of the approaches implemented in this paper on the DRC data set at hand, which should help practitioners to unbox the R code provided.

\paragraph{Raw data}

To mitigate problems due to the high dimensionality and the strong correlation of the data, for every DRC observation consisting of 238 equally spaced points, we chose to keep only 19 points, one each 13, and applied clustering methods on these sliced DRCs. 
When using model-based methods, the selection of the optimal number of clusters was based on the optimization of the BIC criterion.
In the other cases, viz. k-means and hierarchical clustering methods, we relied on the $\textsf{R}$ package \texttt{NbClust}\cite{charrad2012nbclust}, which allows the calculation of several indices.
Then, the optimal number of clusters was chosen according to the majority rule, i.e., as that optimizing the largest number of criteria.

\paragraph{Filtering}

Let us consider the case when using B-spline basis, first.
If we choose too many basis functions to represent profiles, we still have the same high dimensionality and correlation problems as in the raw data case.
This can also make computation very slow.
Since this data set is characterized by functions that are relatively smooth, we chose to regularize using a lower number of basis functions and to avoid the penalization of the integrated squared second derivative.
We selected the number of basis functions on the basis of the generalized cross-validation criterion.
In particular, in order to keep the number of basis functions low, after plotting the generalized cross-validation against the number of bases, we selected 12 basis functions as the elbow of the curve.

When using FPCA in the filtering approach, we applied clustering on the functional principal component scores. 
We first obtained smooth functions using B-spline basis expansion, with 100 basis.
Then, we regularized as described in Section \ref{sec_fil} by means of a penalty on the integrated squared second derivative, with smoothing parameters chosen by minimizing the generalized cross-validation criterion.
Thus, we performed functional principal component analysis on the obtained functional data set and retained only the components that explain the 99\% of the total variability in the data.
This allowed in practice to reconstruct original functions with a strong dimension reduction. 
In fact, since profiles in this data set are smooth, only 6 principal components were required.

For both B-spline and FPCA basis, when using the model-based approach, the selection of the optimal number of clusters was based on optimization of the BIC criterion.
In the other cases, viz. k-means and hierarchical clustering methods, we relied on the $\textsf{R}$ package \texttt{NbClust}\cite{charrad2012nbclust}, which allows the calculation of several indices. 
The optimal number of clusters was chosen also in this case according to the majority rule. 

\paragraph{Adaptive}

The \textit{adaptive fclust} method was performed by means of the $\textsf{R}$ package \textit{fclust}, which requires the choice of the number of basis functions $K$, as mentioned in Section \ref{sec_adme}, that was set equal to 5, 10.
Parameters and the number of clusters were set based on the BIC criterion.

For \textit{waveclust}, we relied on the $\textsf{R}$ package \texttt{curvclust}\cite{curvclust}, dedicated to model-based curve clustering.
In particular, the considered models include Functional Clustering Mixed Models (FCMM, i.e., functional clustering with the presence of functional random effects), but also traditional functional clustering model (FCM, without functional random effects). 
Among FCMMs, several structures of the variance of the random effect can be chosen.
In particular, the following alternatives are available, as mentioned in Giacofci et al.\cite{giacofci2013wavelet}: constant, group, scale-position, and group-scale-position dependent.
It is also possible to decide whether to retain all coefficients or to perform individual denoising to keep coefficients which contain individual-specific information, by applying nonlinear wavelet hard thresholding before clustering.
We considered all parameter combinations of the variance structures and chose the model and number of clusters that optimize the BIC criterion.
Model fitting was performed by maximum likelihood method using the EM algorithm and the stochastic EM as initialization method.

When using \textit{funHDDC}, several parameters need to be chosen. 
We considered 0.2, 0.5, and 0.9 as possible values for the threshold  of the Cattell' scree-test used for selecting the group-specific intrinsic dimensions $d_m$. 
Moreover, the following alternative models are available, as described in Bouveyron and Jacques\cite{bouveyron2011model}: $a_{kj}b_{k}Q_kd_k$, $a_{kj}bQ_kd_k$, $a_kb_kQ_kd_k$, $ab_kQ_kd_k$, $a_kbQ_kd_k$, $abQ_kd_k$.
In this paper, we considered all parameter combinations and chose the model and numbers of clusters that optimize the BIC criterion.
Moreover, to avoid local minima, for each parameter combination we repeated the model fitting 20 times and kept the model with the largest log-likelihood.

\paragraph{Distance-based}

In the distance-based approach, we applied the clustering algorithm for each number of clusters by considering the distance in Equation \eqref{eq_dist} with $l=0$, that is the usual $L^2$ distance, then, we kept the model with the best value of the silhouette index.

\end{document}